\documentclass[]{spie}
\usepackage{graphicx}
\usepackage{dcolumn}
\usepackage{bm}
\usepackage{upgreek}
\usepackage{float}
\title{Towards on-chip generation, routing and detection of non-classical light} 

\author{F. Flassig\supit{1}, M. Kaniber\supit{1}, G. Reithmaier\supit{1}, K. M\"uller\supit{2}, A. Andrejew\supit{1}, R. Gross\supit{3,4}, J. Vu\c ckovic\supit{2} and J. J. Finley\supit{1,4}
\skiplinehalf
\supit{1}Walter Schottky Institut and Physik Department, Technische Universit\"at M\"unchen, Am Coulombwall 4, 85748 Garching, Germany; \\
\supit{2}E. L. Ginzton Laboratory, Stanford University, Stanford, CA 94305, USA; \\
\supit{3}Walther Meißner Institut and Physik Department, Technische Universit\"at M\"unchen, Walther-Meißner-Straße 8, 85748 Garching, Germany; \\
\supit{3}Nanosystems Initiative Munich (NIM), Schellingstraße 4, 80799 M\"unchen, Germany
}
\authorinfo{Corresponding author: michael.kaniber@wsi.tum.de; phone +49 89 289 12782}

\date{\today}

\begin{document}
\maketitle 
\begin{abstract}
We fabricate an integrated photonic circuit with emitter, waveguide and detector on one chip, based on a hybrid superconductor-semiconductor system. We detect photoluminescence from self-assembled InGaAs quantum dots \textit{on-chip} using NbN superconducting nanowire single photon detectors. Using the fast temporal response of these detectors we perform time-resolved studies of non-resonantly excited quantum dots. By introducing a temporal filtering to the signal, we are able to resonantly excite the quantum dot and detect its resonance fluorescence on-chip with the integrated superconducting single photon detector.

\end{abstract}

\keywords{Superconducting nanowire single photon detectors, quantum dots, integrated photonics}
\section{Introduction}
%
%
Semiconductor based photonic information technologies are rapidly being pushed to the quantum limit where single photon states can be generated, manipulated and exploited in nanoscale optical circuits\cite{Politi09,Matthews09}. Over recent years quantum dots (QDs) embedded in such semiconductor systems have been shown to be excellent sources of quantum light\cite{Flagg09,Gao12,He13} and have demonstrated their suitability for use as a gain medium in QD lasers\cite{Ellis11,Ledentsov98,Ellis07}. Superconducting single photon detectors (SSPDs), on the other side, have emerged as highly promising single photon detectors, showing very high detection efficiencies \cite{Marsili08,Hu09,Kerman07,Reithmaier13}, low dark count rates\cite{Kitaygorsky07}, sensitivity from the visible to the IR\cite{Hofherr10} and picosecond timing resolution\cite{Goltsman05,Najafi12}. Building up on recent progress in this field\cite{Pernice12,Sprengers11,Sahin13}, we developed highly efficient\cite{Marsili08,Kerman07} NbN-SSPDs on GaAs\cite{Reithmaier13} and demonstrated the monolithic integration of InGaAs QDs as single photon emitters together with waveguides and detectors on a single chip\cite{Reithmaier13SciRep}.
In this work, we present the fabrication of hybrid superconductor-semiconductor systems with self-assembled InGaAs QDs as quantum emitters embedded in ridge GaAs waveguides and NbN-SSPDs on one chip, forming a simplistic integrated photonic circuit. We perform optical investigations of the QDs embedded in the ridge waveguide and exploited the intrinsically fast temporal response of the SSPDs to study the time-resolved photoluminescence from QDs in situ. We introduce measures that faciliate the elimination of scattered background laser light in the detector and, by additionally including a temporal filtering of the recorded sognal, we detect QD resonance fluorescence with a transition linewidth of 20 $\pm$ 3 $\upmu$eV on-chip using the integrated SSPD.

\section{SSPD operation principle: Electro-thermal model}
\begin{figure}[htb]
	\centering
\includegraphics[width=0.95\columnwidth]{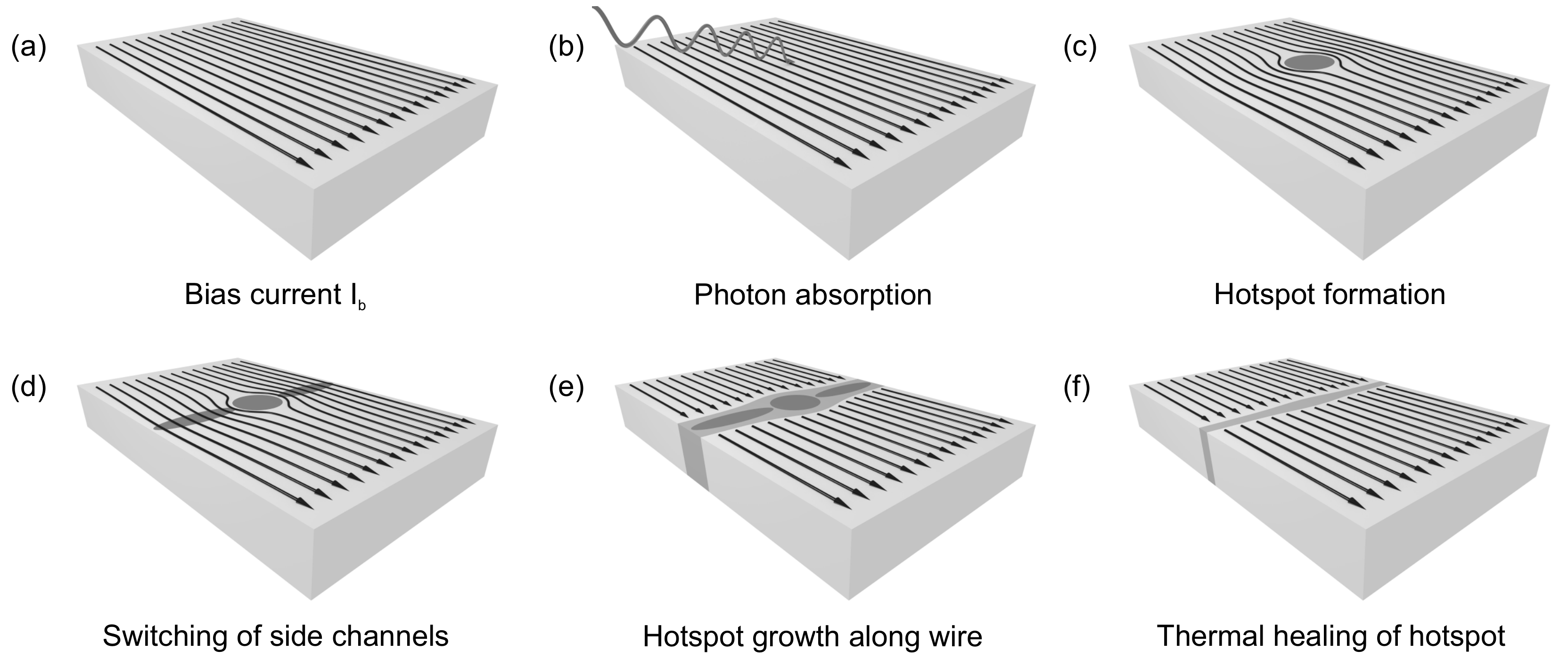}
\caption{\label{fig1} Schematic of hotspot formation process: (a) A superconducting nanowire is DC biased close to its critical current. (b) The absorption of a photon with energy $\hbar \omega$ leads to (c) the formation of a local region with suppressed superconductivity, redistributing the bias current around this \textit{hotspot}. (d) The increased current density in the side channels exceeds the critical current density, forming a resistive barrier across the whole width. (e) Joule heating of the resistive barrier causes a hotspot growth along the wire. (f) The current is quenched and the hotspot heals out via thermal diffusion to the substrate.}
\end{figure}

%
%
Our superconducting single-photon detectors consist of a DC reactive magnetron sputtered 10 nm thick NbN film being patterned into narrow nanowires, each of them with a width of 100 $\pm$ 5 nm. In our case the device is realized on a GaAs substrate, as it is described in more detail in the next section. For operation, the device is kept well below its superconducting transition temperature $T_C$, typically at temperatures lower than $T_C/2$\cite{Kit08}. The basic operating principle of an SSPD is schematically illustrated in figure \ref{fig1}. A bias current $I_b$ is sent through the detector which is $\sim 95\%$ of the critical current $I_C$, as depicted in figure \ref{fig1} (a). Upon absorption of a photon in the wire (b), the absorbed photon energy leads to a locally increased electron temperature and suppression of the superconductivity\cite{Ili98, Goltsman05}, a so-called \textit{hotspot} is formed (c). Here, the arrows represent the superconducting bias current being expelled into the side channels of the nanowire. The current density in the side channels exceeds the
critical current density $J_C$ causing the wire to become normal conducting along the entire width
of the stripe (d). As a result of the current flowing through the detector, Joule heating leads to a hotspot growth along the wire (e), while at the same time the current is quenched, which allows the hotspot to thermally heal out. 
Due to the additional resistivity, the current can be redirected into an output circuit as presented in figure \ref{fig3} (a), allowing for the detection of a single voltage pulse for each photon being absorbed within the superconducting detector.

\section{Sample fabrication and layout}
\begin{figure}[t]
	\centering
\includegraphics[width=0.95\columnwidth]{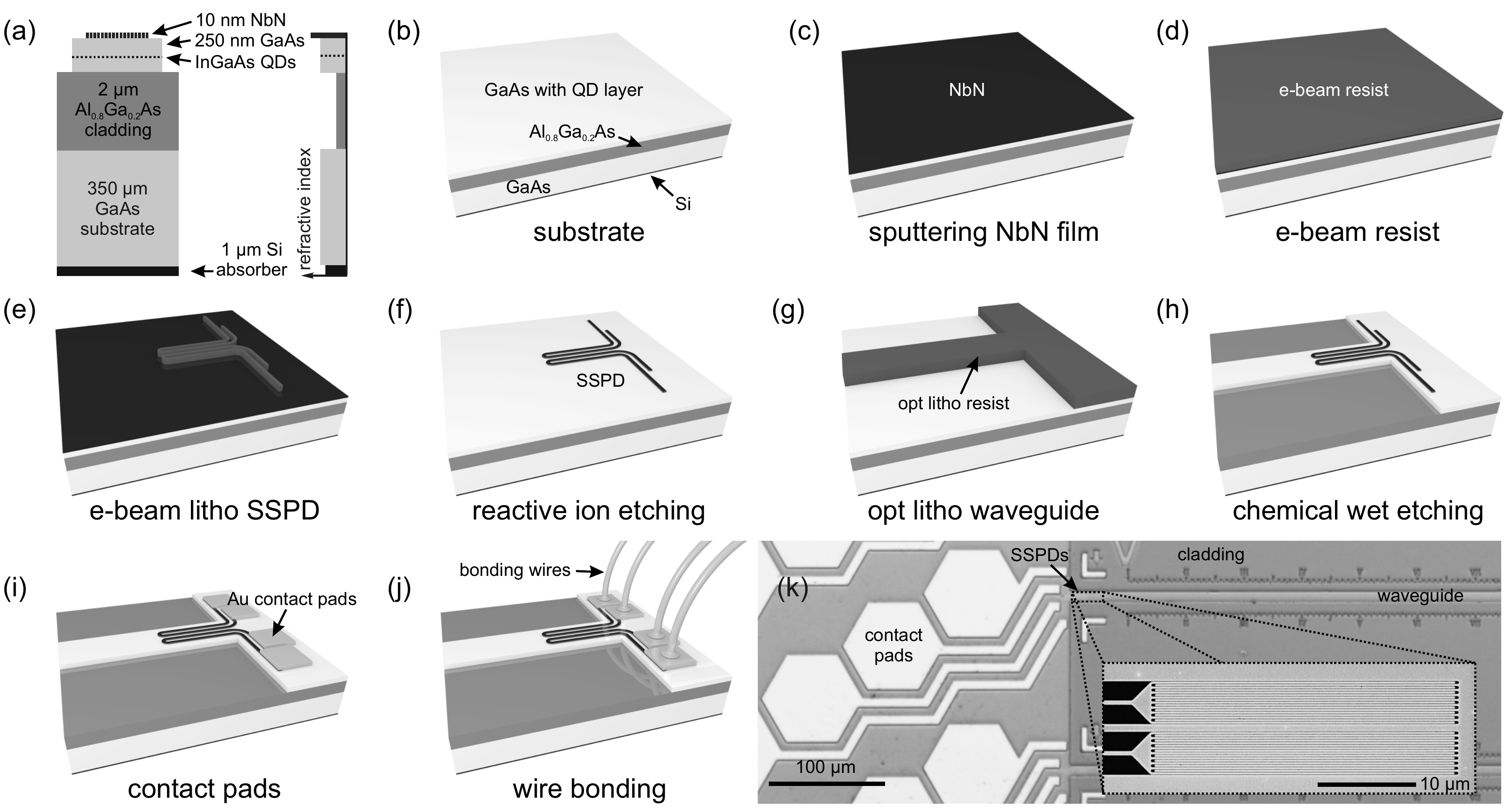}
\caption{\label{fig2} (a) Schematic cross section through the sample structure on the left side with the according refractive indices on the right side. (b) - (j) Fabrication steps for producing waveguide-coupled SSPDs. (k) Microscope image of fabricated SSPDs coupled to GaAs ridge waveguides. The SEM image in the inset depicts the pair of waveguide-coupled SSPDs. }
\end{figure}

To fabricate our hybrid superconductor-semiconductor systems of SSPDs on III-V semiconductor ridge waveguides, we use a combination of DC reactive magnetron sputtering, electron beam lithography and reactive ion etching.
We start our fabrication with an In$_{0.4}$Ga$_{0.6}$As/GaAs QD sample grown by standard molecular beam epitaxy. On top of the nominally undoped GaAs (100) substrate a 2 $\upmu$m thick layer of Al$_{0.8}$Ga$_{0.2}$As is grown, followed by a 250 nm thick layer of GaAs with self-assembled InGaAs QDs in its center. This layer structure is schematically shown in figure \ref{fig2} (a). Due to the refractive index contrast between GaAs and AlGaAs, the 250 nm thick layer of GaAs acts as a waveguide. When positioning a NbN detector on top of it, light inside the waveguide will evanescently couple to the detector, leading to an efficient absorption of light in the detector\cite{Reithmaier13SciRep}. 
The QDs are grown in the Stranski–Krastanov growth mode at a temperature of 560$^\circ$ C by depositing 3.5 monolayers of nominally x = 50$\%$ In$_x$Ga$_{1-x}$As, giving hemispherically shaped self-assembled QDs with 25 nm in diameter and 5 nm in height with an areal density of 2-3 $\upmu$m$^2$ \cite{PHDMue13}.
The backside of the sample is mechanically polished and coated with a 1 $\upmu$m thick layer of Si to absorb stray light with energies below the GaAs bandgap. 
On top of the substrate, a 10 nm thin layer of high-quality superconducting NbN evaporated by DC reactive magnetron sputtering\cite{Reithmaier13}, as shown in figure \ref{fig2} (c).
This film is spin coated with a negative tone electron beam resist (AR-N 7520, Allresist) at 4000 rpm for 40 s, then softbaked for 300 s at 85$^\circ$ C (d). The detectors are patterned using electron beam lithography. After developing the exposed resist for 110 s (AR 300-47, 4:1, Allresist) (e), the sample is reactively etched in a SF$_6$/C$_4$F$_8$ plasma to transfer the detector structures onto the NbN film (f).
To form the ridge GaAs waveguides, a positive tone optical resist (S1818, Microposit) is spin-coated onto the sample at 5000 rpm for 40 s and softbaked for 300 s at 90$^\circ$. The waveguide structure is oriented with respect to the SSPDs and written into the resist by optical lithography and the resist is developed for 25 s (AZ351B, 1:4, AZ Electronic Materials) (g). The GaAs layer is wet-chemically etched in a highly diluted citric acid hydrogen peroxide solution (1 g CA : 100 ml H$_2$O : 100 ml H$_2$O$_2$) for 30 min at 35$^\circ$ C to produce smooth waveguide sidewalls necessary for low propagation losses. The remaining resist is removed with acetone (h).
The contact pad structures for the detectors are fabricated similarly to the waveguides. The sample is spin coated with a positive tone optical resist (S1818, 40 s, 5000 rpm). The contact pad structure is written into the resist and the resist is developed (AZ351B, 1:4, 25 s). The sample is dipped in 10$\%$ HCl for 30 s to remove the natural surface oxide layer of the GaAs. A 10 nm thick adhesive layer of Ti, followed by a 100 nm thick layer of Au are evaporated onto the sample. After removing the resist mask, the contact pads remain on the sample (i). In a final step, aluminum wires are ultrasonically bonded to the pads and connected to the read-out electronics (j).
A tyüical microscope image of a fabricated sample is shown in figure \ref{fig2} (k). At both ends of a nominally 2.6 $\upmu$m long, 20 $\upmu$m wide multimode ridge waveguide a pair of SSPDs, each consisting of 18$\times$, 10 nm thick, 100 nm wide and 23 $\upmu$m long NbN nanowires, as depicted in the SEM inset in figure \ref{fig2} (k). These detectors provide a near-unity absorption efficiency of light inside the waveguide\cite{Reithmaier13SciRep}. The hexagonal contact pads have an inner diameter of 150 $\upmu$m to provide enough space for tolerances in the bonding wire positioning.

\section{Experimental Setup}
\begin{figure}[t]
	\centering
\includegraphics[width=0.95\columnwidth]{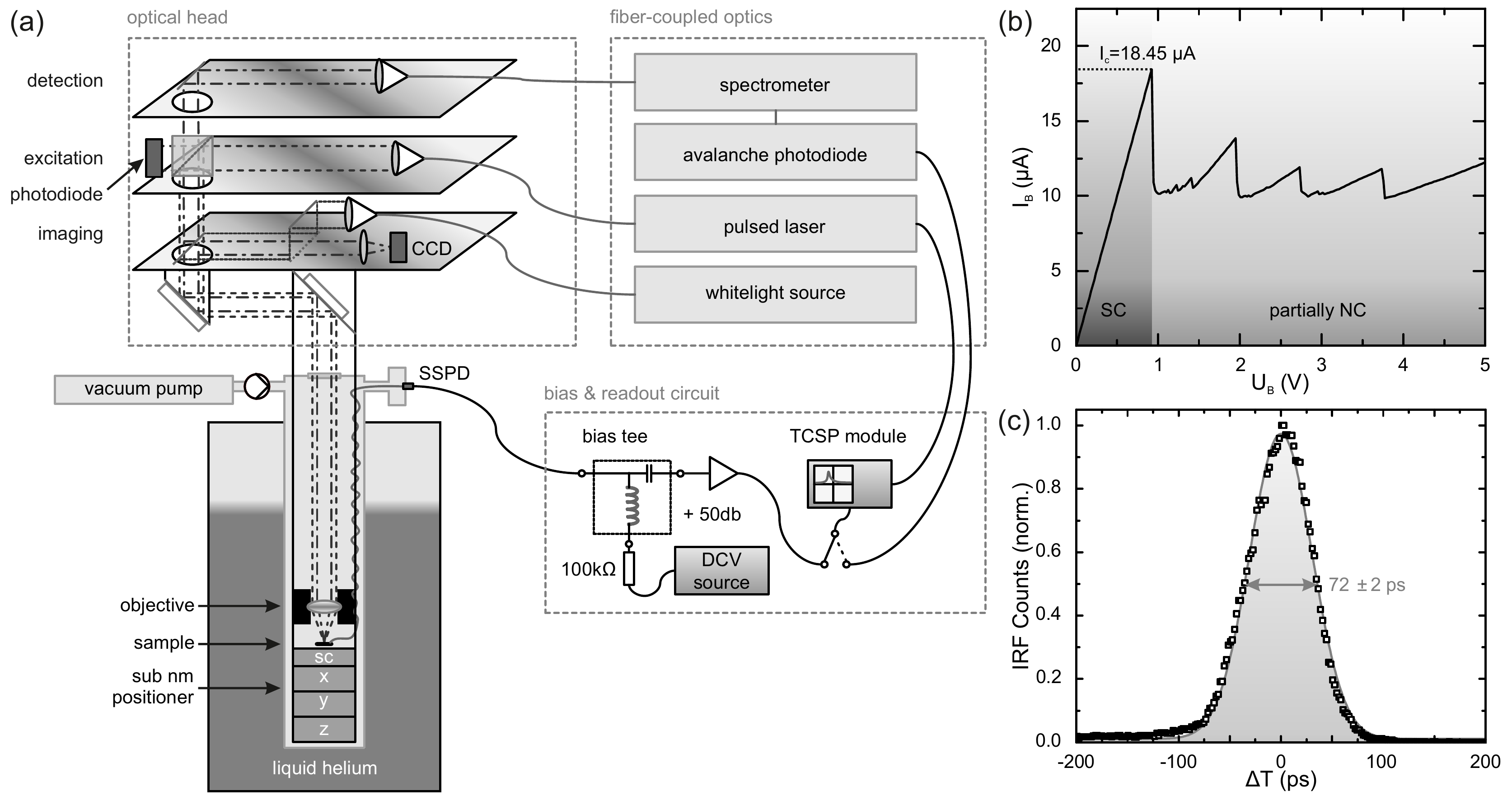}
\caption{\label{fig3} (a) Schematic of the experimental setup and the electrical read-out circuit. (b) Current-voltage characteristic of a 10 nm thick SSPD at 4.2 K. The dark gray shaded region marks the fully superconducting regime in which the detector is operated. In the light gray shaded region the detector is partially normal conducting. (c) Temporal response of the SSPD for illumination with a sub-70 ps pulsed laser diode.}
\end{figure}

%
A stable environmental temperature below $T_C/2$ is required to allow for efficient and reliable operation of SSPD detectors \cite{Kit08}. With critical temperatures of the NbN film well above 10 K\cite{Reithmaier13}, a dipstick cryostat, as shown in figure \ref{fig3} (a), provides a long-term stable environment at liquid helium temperature combined with a micro-photoluminescence microscope. 
The stick is pumped to pressures $\leq 1 \cdot 10^{-5}$ before it is filled with helium acting as an exchange gas. Nano-positioners inside the stick allow for positioning the sample with respect to the laser focus. Optical access is provided via an objective and a window to the optical head which consists of three layers that are schematically depicted in the top left of figure \ref{fig3} (a). The lowest level contains a charge coupled device (CCD) and a white light incoupling for imaging the sample surface. The middle layer contains the excitation path and is fiber-coupled to the excitation laser. A photodiode is used to determine the incident laser power. Via a 50:50 unpolarized beamsplitter the upper level is connected, which contains the detection path and is fiber-coupled to a spectrometer with both a liquid nitrogen cooled CCD and a single photon avalanche photodiode (SPAD) used to measure spectrally and temporally resolved signals. The integrated SSPDs can be operated via a bias tee. A constant voltage source and a 100 k$\Omega$ resistor provide the bias current for the SSPD. At the third side of the bias tee which is blocked for the DC current via a capacitor different read-out electronics can be connected. Here, two amplifiers with overall +50 dB amplification and a time-correlated single photon counting (TCSPC) module with a temporal resolution of 150 ps are used. 
Figure \ref{fig3} (b) shows the current voltage characteristics of a 10 nm thick SSPD at 4.2 K when increasing the bias voltage. Below a critical voltage $U_C$, corresponding to a critical current $I_C = 18.45 \mu$A for this particular device, the current shows a linear dependence, corresponding to the 100 k$\Omega$ resistor. At $I_C$, an abrupt decrease of the current is observed, which repeats several times for increasing bias voltages. This behavior can be explained by two different operational regimes. For $U < U_C$ (dark gray shaded region), the whole detector is superconducting. When the current reaches the critical current, superconductivity breaks down in the nanowire with the smallest cross section, causing an additional resistivity (light gray shaded region). This process repeats for the other wires until the critical current is exceeded in all wires and the whole device is normal conducting. To provide efficient operation of the SSPD, a bias current close to the critical current, $I_b \sim 0.95 I_C$, is chosen.
Attaching a 20 GHz band-width sampling oscilloscope to the read-out arm of the bias tee allows us to measure the instrument response function (IRF) of our SSPDs. Figure \ref{fig3} (c) shows the timing jitter of a typical detector as black squares when illuminating the detector directly with a 1.302 eV pulsed laser diode with $\leq$ 70 ps pulse duration. A Gaussian fit, shown in gray, reveals a full width half maximum of 72 $\pm$ 2 ps, most likely limited by the laser pulse duration used. SSPDs of similar geometries were reported to exhibit values as low as 18 ps \cite{Goltsman05,Pernice12}.

\begin{figure}[htb]
	\centering
\includegraphics[width=0.95\columnwidth]{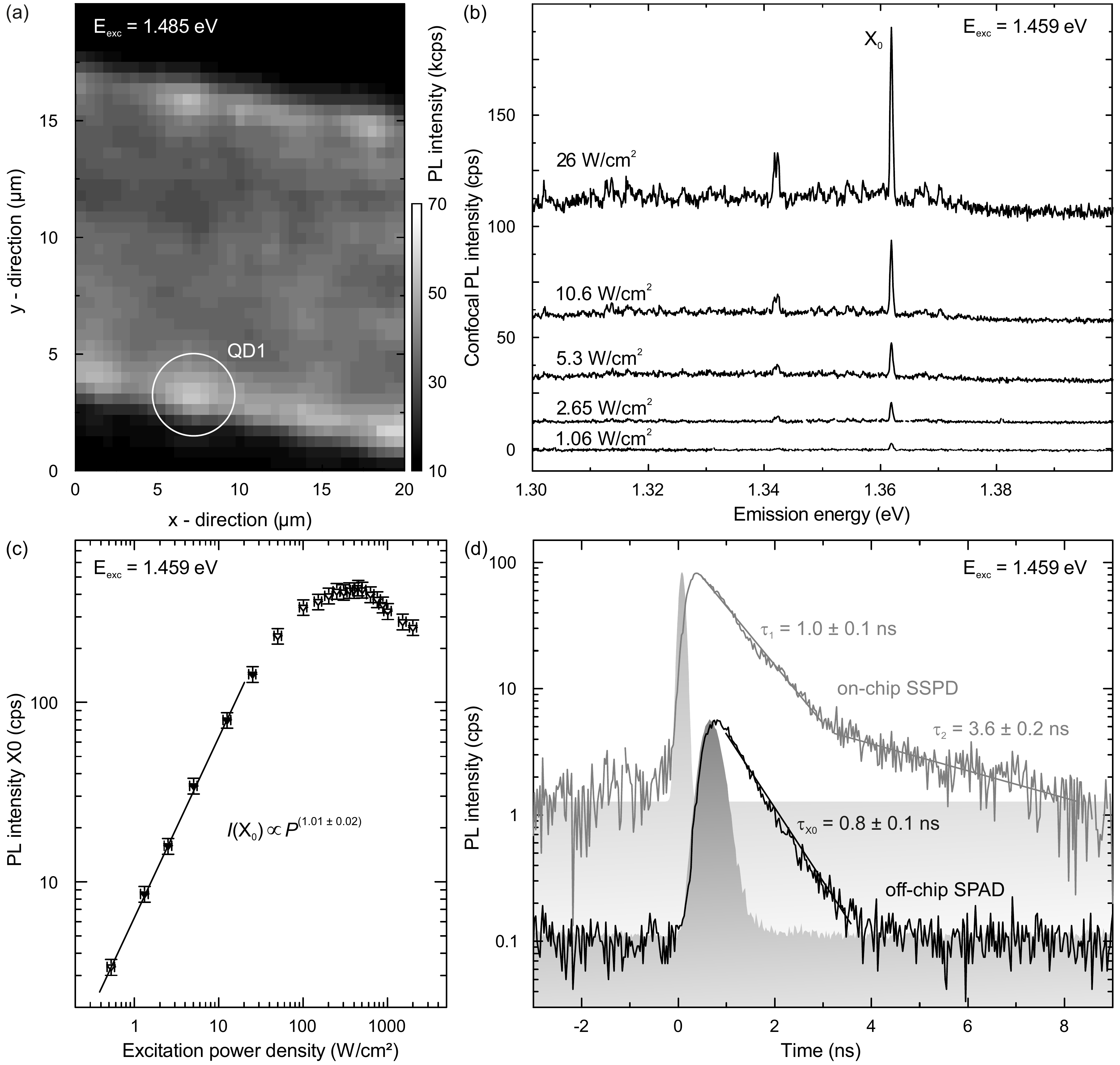}
\caption{\label{fig4} (a) On-chip SSPD signal for a spatial scan of the excitation spot over the waveguide for an excitation energy of $E_{exc}$ = 1.485 eV. The white circle marks the position of the excitation spot used to obtain data shown in panels (b) - (d). (b) Low power PL spectra of QD1 for an excitation energy of $E_{exc}$ = 1.459 eV. (c) Power dependent PL intensity of $X_0$ in double-logarithmic plot for this $E_{exc}$. (d) Time resolved signal for an excitation power density of $25.9 \pm 0.6$ Wcm$^{-2}$. The gray curve shows the spectrally integrated signal on the on-chip SSPD, while the black curve shows the spectrally filtered signal of $X_0$ on an off-chip avalanche photodiode. The gray shaded regions mark the corresponding instrument response functions.}
\end{figure}
\section{Time-resolved QD luminescence}
%
In this section we present optical investigations of a QD embedded in a ridge GaAs waveguide on a similar sample, as shown in figure \ref{fig2} (k), containing one detector at each end of a 350 $\upmu$m long waveguide. As the integrated SSPDs on a ridge waveguide provide no intrinsic spectral selectivity, a locally isolated QD is necessary to ensure that only light stemming from this dot is detected in the SSPD. Confocal photoluminescence (PL) spectroscopy hereby enables us to study the spectral response of the QD. 
Figure \ref{fig4} (a) shows the on-chip SSPD signal for a spatial scan of the excitation laser spot over a waveguide segment for a cw-laser diode with $E_{exc}$ = 1.485 eV. Excitation on the waveguide, which is roughly oriented along the x-direction,  produces a nearly one order of magnitude higher signal than excitation on the Al$_{0.8}$Ga$_{0.2}$As buffer layer. In the following, we focus on one bright spot at the edge of the waveguide, as marked \textit{QD1} in figure \ref{fig4} (a). 
Throughout the following, we employ a sub-ps pulsed laser with an energy of $E_{exc} \sim 1.459 eV$ within the wetting layer energy of the QDs to enable time-resolved measurements.
Figure \ref{fig4} (b) shows PL spectra of an isolated dot with one predominant, spectrally isolated emission line $X_0$, for low excitation power densities varying from 1.06 Wcm$^{-2}$ to 26 Wcm$^{-2}$. The additional weak emission lines at lower energies either stem from the emission of nearby QDs or charged excitonic transitions of this QD.  
The power-dependency of $X_0$ is shown in figure \ref{fig4} (c). For low excitation power densities, a clearly linear power dependency is apparent with a power law of $I \propto P^{(1.01 \pm 0.02)}$. For higher excitation power densities, the PL intensity saturates and decreases, therefore, we attribute this line to a single exciton transition\cite{Finley01}.
We continue by spectroscopically filtering the $X_0$ transition via a monochromator and measuring the time-resolved PL signal at a low excitation power density of $25.9 \pm 0.6$ Wcm$^{-2}$ on a single photon avalanche photodiode connected to a TCSPC module. The recorded time trace is shown as a black line in figure \ref{fig4} (d) with the according IRF shaded in dark gray. While the rise time of the signal is limited by the timing performance of the detection electronics, a clear mono-exponential decay is apparent with a lifetime $\tau_{QD1} = 0.8 \pm 0.1$ ns. This corresponds to the lifetime of the $X_0$ transition and compares very well to the known spontaneous emission lifetimes of such InGaAs QDs \cite{Rao07,Schwoob05}.
We repeat this measurement using the on-chip integrated SSPD. The signal recorded on-chip by the SSPD, as shown by the gray curve in figure \ref{fig4} (d), is over one order of magnitude higher than the signal recorded off-chip by the SPAD, an observation attributed to the efficient coupling of QD PL to the waveguide and of waveguide photons to the detector \cite{Reithmaier13SciRep}. In addition, due to the missing spectral selectivity of the on-chip SSPD, it collects also luminescence signal from other QDs. 
The SSPD IRF of this configuration is shaded in light gray. Due to the high system detection efficiency and the fast timing resolution, the integrated SSPD allow for time-resolved studies of QD decay dynamics\cite{Reithmaier14}.
The gray curve in figure \ref{fig4} (d) reveals a bi-exponential decay with a bright fast component with $\tau_{1} = 1.0 \pm 0.1$ ns and a slower component, $\tau_{2} = 3.6 \pm 0.2$ ns, with less intensity. While the fast decay can be attributed to the $X_0$ transition, as it is in accordance with the off-chip recorded decay of $X_0$ within the measurement uncertainty, the slower decay is attributed to other transitions.

\begin{figure}[H]
	\centering
\includegraphics[width=0.95\columnwidth]{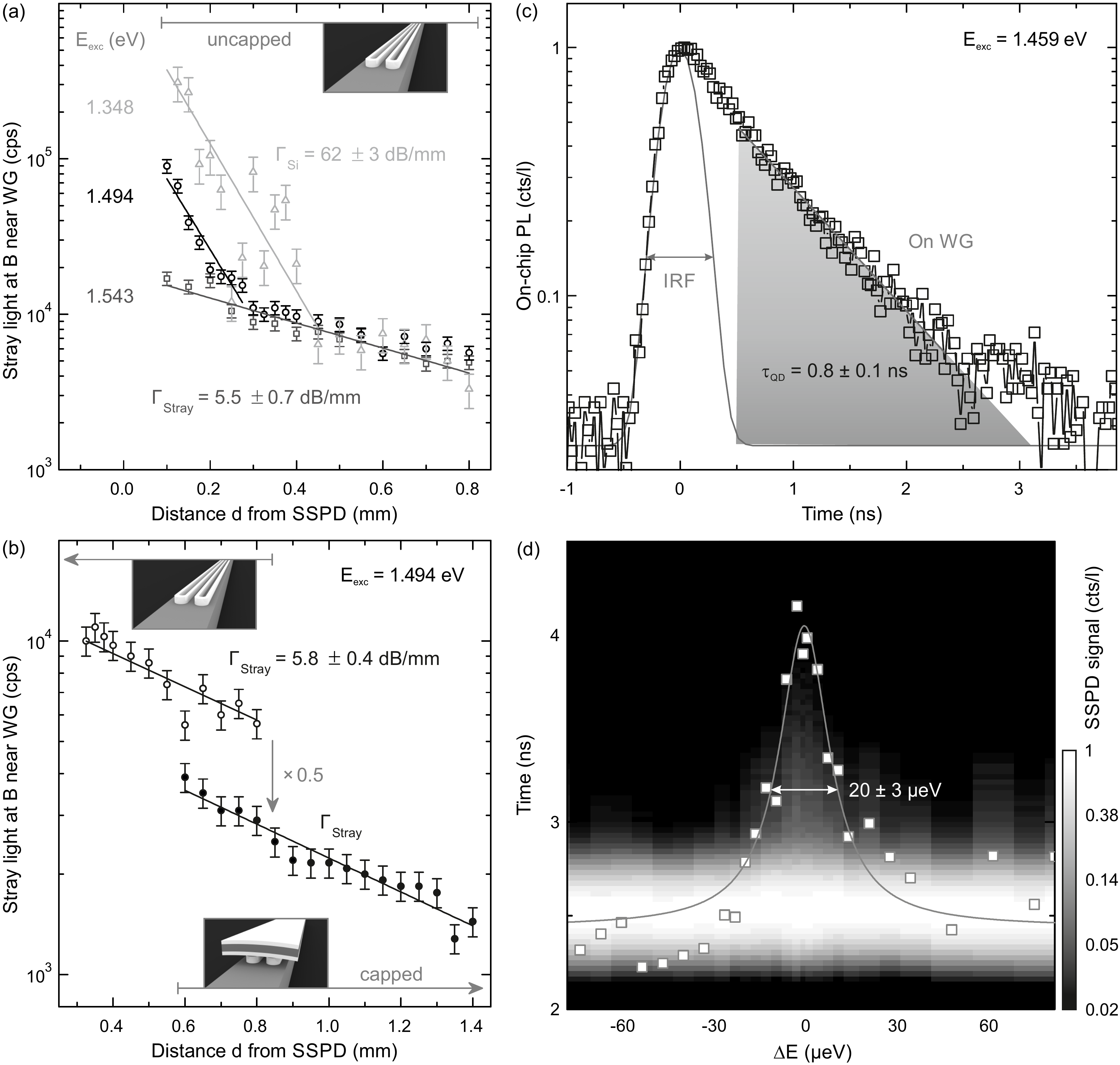}
\caption{\label{fig5} (a) On-chip SSPD \textit{stray light} signal for illumination 20 $\upmu$m besides the waveguide. The data is shown as a function of distance $d$ from the detector for three different illumination energies. For illumination at an energy below the GaAs badgap (light gray and black data points), a fast and a slow exponential decay are observed, whereas for 1.543 eV (dark gray) only the slow component is apparent. The corresponding propagation lengths are 70 $\pm$ 4 $\upmu$m for $\Gamma_{Si}$ and 796 $\pm$ 96 $\upmu$m for $\Gamma_{Stray}$, respectively. (b) Measured off-waveguide stray light signal as a function of distance $d$ comparing an uncapped detector (open symbols) with a teflon-aluminum-teflon capped detector (full symbols). In both cases the same propagation length of 752 $\pm$ 55 $\upmu$m can be extracted. (c) On-chip time-resolved SSPD signal of multiple QDs for excitation energies in the wetting layer. The IRF is shown in gray and corresponds to the signal obtained for direct illumination of the detector with a 300 ps pulsed diode laser. To temporally filter out laser stray light contributions to the SSPD signal, only the signal in the gray shaded region is considered which solely stems from QD emission. (d) Contour plot: Time transients of the resonance fluorescence signal as a function of energy detuning, measured with a 300 ps diode laser. The data is color coded in counts per detected laser photon (cts/l). White data points: Time-integrated RF signal for a time window from 0.5 ns to 5 ns after the signal peak intensity, as indicated by the gray shaded region in panel (c). A Lorentzian line shape (gray solid line) with a FWHM of 20 $\pm$ 3 $\upmu$eV is fitted to these data points.}
\end{figure}

In order to detect light from only a single QD transition instead of from multiple QDs, we move from a non-resonant to a resonant excitation of a single QD to provide spectral selictivity in the excitation channel. Hereby, on a GaAs substrate with an unpolished and unprocessed sample backside, a strong laser background signal in the on-chip detector becomes apparent. This signal mainly stems from laser light reflection at the sample backside. To eliminate this \textit{stray light}, we polish the sample backside and coat it with a layer of amorphous Si, as schematically depicted in figure \ref{fig2} (a). The effect of the Si-coated backside is depicted in figure \ref{fig5} (a), that shows the on-chip detected signal plotted as a function of the excitation spot distance from the SSPD at three different excitation energies for excitation spot positions 20 $\upmu$m besides the waveguide. By exciting off-waveguide on the buffer layer, no QDs are excited and the detected signal only reflects laser stray light that gets absorbed by the detector. For excitation energies below the GaAs bandgap (light gray and black curve), two exponential decays are observed with fast and slow time constants, with propagation lengths of 70 $\pm$ 4 $\upmu$m for $\Gamma_{Si}$ and 796 $\pm$ 96 $\upmu$m for $\Gamma_{Stray}$, respectively. For an excitation energy above the GaAs bandgap, only the slow component is observed. Therefore, we attribute $\Gamma_{Si}$ to laser light scattered at the sample backside and $\Gamma_{Stray}$ to laser light reaching the detector from the environment. By increasing the waveguide length to $\geq$ 500 $\upmu$m, light scattered at the backside can be effectively suppressed. To eliminate the remaining stray light, we add a capping layer to the detector, which consists of an opaque teflon-aluminum-teflon multilayer structure. Figure \ref{fig5} (b) shows the effect of capping the detector for an excitation energy within the QD wetting layer as a function of excitation distance from the detector. Both the uncapped detector (open symbols) as well as the capped detector (full symbols) show a slow exponential decay, which compares well to propagation losses of similar ridge GaAs waveguides \cite{Reithmaier14}.

\newpage
\noindent While the capping further reduces stray light by a factor of 2$\times$, it cannot completely eliminate the stray light and we, therefore, attribute the remaining stray light to residual background radiation reaching the SSPD or to surface luminescence created at the Al$_{0.8}$Ga$_{0.2}$As surface. 
Therefore, we introduce in addition a temporal filtering to the on-chip detected time-resolved signal to completely suppress the background signal. Thereby, we use a 300 ps pulsed diode laser. The IRF of this laser, when directly illuminating the SSPD, is displayed as a gray Gaussian curve. By integrating the recorded signal in a time window after the laser pulse, as indicated by the gray shaded rectangle, we can eliminate the laser stray light from the SSPD signal.
In figure \ref{fig5} (d), we tune the excitation laser over a quantum dot transition and record the time-resolved on-chip luminescence signal. The contour plot shows these time transients as a function of energy detuning from a QD transition. The horizontal white stripe corresponds to the laser stray light, which is independent of the excitation energy. At zero energy detuning, a peak is apparent. By integrating these transients as indicated by the gray shaded area in figure \ref{fig5} (c) in a time window from 0.5 ns to 5 ns, we extract the RF signal, which is plotted as white squares in figure \ref{fig5} (d). By fitting a Lorentzian line shape to these points, we extract a linewidth of 20 $\pm$ 3 $\upmu$eV of this transition, in good agreement with values reported for off-chip detected RF from QDs embedded in a free standing single-mode ridge waveguide \cite{Makhonin14}.

\section{Conclusion}
In summary, we developed an integrated photonic circuit with quantum emitters, ridge waveguides and single photon detectors on one chip based on a hybrid superconductor-semiconductor structure. By using a combination of DC reactive magnetron sputtering, electron beam lithography and reactive ion etching we fabricated high-quality SSPDs on top of ridge GaAs waveguides. To allow for a stable operation of the SSPDs and optical access to the QDs, we use a dipstick cryostat with an integrated micro-photoluminescence microscope. With this setup we probed the temporal response of MBE grown self assembled InGaAs QDs embedded in the waveguide in-situ with the integrated SSPDs for non-resonant excitation. To overcome the missing spectral selictivity of our photonic circuit, we resonantly adressed a single QD transition. We were able to suppress scattered laser light in the detector by employing a combination of various techniques, namely a polished sample backside coated with a layer of Si in combination with $\geq 1$ mm long waveguides to avoid light scattered at the sample backside, a capping of the detector to avoid stray light from the top, and a temporal filtering of the recorded signal. With these measures we were able to resonantly excite a QD transition and detect its resonance fluorescence signal on-chip using the integratedm SSPDs, hereby revealing a transition linewidth of 20 $\pm$ 3 $\upmu$eV. 

\acknowledgments
We gratefully acknowledge D. Sahin, A. Fiore (TU Eindhoven) and K. Berggren, F. Najafi (MIT) and R. Hadfield (University of Glasgow) for useful discussions and the BMBF for financial support via QuaHL-Rep, project number 01BQ1036, and Q.com via project number 16KIS0110, the EU via the integrated project SOLID and the DFG via SFB 631-B3 and the ARO (grant W911NF-13-1-0309).

\bibliography{References}
\bibliographystyle{spiebib}   

\end{document}